\documentclass[10pt]{article}

\usepackage{graphics}
\usepackage{graphicx}

\usepackage[a4paper, left=35mm,right=35mm,top=34mm,bottom=34mm]{geometry}
\usepackage[utf8]{inputenc}
\usepackage[T1]{fontenc}
\usepackage[english]{babel}

\usepackage{enumerate}
\usepackage{graphicx}
\usepackage{hyperref}
\hypersetup{
    colorlinks=true,
    linkcolor=blue,
    filecolor=magenta,      
    urlcolor=cyan,
}

\usepackage{mathtools,amsthm,amssymb,amsfonts}
\usepackage{algorithm}
\usepackage{algorithmic}
\makeatother
\theoremstyle{plain}

\theoremstyle{definition}

\theoremstyle{remark}

\newtheorem{alg}{Algorithm}

\usepackage{caption} 
\usepackage{fancyhdr}

\newcommand{\abs}[1]{\left\lvert#1\right\lvert}

\author{
  {\normalsize Qinmeng Zou}\thanks{CentraleSup\'elec, Universit\'e Paris-Saclay, France.}
  \and
  {\normalsize Guillaume Gbikpi-Benissan}\thanks{CentraleSup\'elec, Universit\'e Paris-Saclay, France.}
	\and
  {\normalsize Fr\'ed\'eric Magoul\`es}\thanks{CentraleSup\'elec, Universit\'e Paris-Saclay, France
    (correspondence, frederic.magoules@hotmail.com).}
		}
\title{Asynchronous Parareal Algorithm Applied to European Option Pricing}
\date{}

\begin{document}
\maketitle
\thispagestyle{fancy}

\begin{abstract}
\noindent Asynchronous iterations arise naturally in parallel computing if one wants to solve large problems with a minimization of the idle times.
This paper presents an original model of asynchronous iterations for a time-domain decomposition method, namely the parareal method.
The asynchronous parareal algorithm is here applied to European option pricing, and numerical experiments performed on a parallel supercomputer, illustrate the performance and efficiency of this new method.
\end{abstract}


\section{Introduction}

An option is a contract that gives the right to buy or sell the underlying asset at a specific price within a period of time, which gives the buyer the right, but not the obligation to perform the transaction.
Option pricing is a crucial target in financial decision-making.
In the practical side, however, it is also one of the most complicated problems in the mathematical domain of financial engineering.
The breakthrough came in 1973, when Black and Scholes proposed the original Black-Scholes equation \cite{Black1973}, which is the first complete European option pricing model.
With an alternative method, Merton derived the formula and generalized it in important ways \cite{Merton1973, Merton1974}.
The Black-Scholes model has a huge influence to the financial market and drives an unexpected prosperity in the trading of derivatives.

With the advent of parallel computers, many complicated problems can be solved efficiently by the improving processing power and new algorithms.
An example is the asynchronous iterative scheme.
The pioneering research of such approach dates back to 1969 by Chazan and Miranker \cite{Chazan1969} that arises in the face of communication delays.
Their works have been extended to the nonlinear cases in \cite{Miellou1975, Baudet1978, ElTarazi1982} based on the notion of contraction.
A well known systematic formalizations is contained in the book of Bertsekas and Tsitsiklis \cite{Bertsekas1989a}.
Furthermore, two-stage theories were proposed in the 1990s by Frommer and Szyld
\cite{Frommer1994, Frommer1998}.
We refer the reader to \cite{Frommer2000, Kollias2011} as surveys on asynchronous iterative methods.
Recently, asynchronous iterations have been extended to efficient domain decomposition methods in space \cite{Toselli2005}, including sub-structuring methods \cite{Magoules2016a}, optimized Schwarz method \cite{Magoules2017a}.
Another issue is the extension to domain decomposition methods in time.
In this paper, we are interested in the asynchronous parareal scheme, which will be explored in
Section \ref{sec:2}.
In Section \ref{sec:3} we present the transformation from the original Black-Scholes equation to an appropriate form which applies to our asynchronous solver.
Section \ref{sec:4} is devoted to the experimental results.
Finally, concluding remarks and future directions are presented in Section \ref{sec:5}.

\section{Asynchronous Parareal Algorithm}
\label{sec:2}

\subsection{Classical Parareal Scheme}

The parareal algorithm is a method to solve the time-dependent partial differential equations in parallel, first presented by Lions, Maday and Turinici in 2001 \cite{Lions2001}.
The ideas of parareal are relevant to both multiple shooting method and predictor-corrector scheme, see the historical remarks in \cite{Gander2007} whereby a comprehensive analysis of parareal has been mathematically studied.

Given a second-order linear elliptic operator $\mathcal{L}$, consider the following time-dependent problem
\begin{equation}
\label{eq:pde1}
\begin{cases}
\frac{\partial u}{\partial t} + \mathcal{L}u = f, & t \in [0, T], \\
u = u_0, & t = 0,
\end{cases}
\end{equation}
where the boundary conditions are unnecessary to be mentioned.
By introducing a function $\lambda_n$, the above mathematical description can be decomposed into $N$ sequential problems, according to the rule $T_n = n\Delta T$ with $n = 0, \dots, N$.
We now rewrite our problem (\ref{eq:pde1}) as
\begin{equation}
\label{eq:pde2}
\begin{cases}
\frac{\partial u_n}{\partial t} + \mathcal{L}u_n = f_n, & t \in [T_n, T_{n+1}], \\
u_n = \lambda_n, & t = T_n,
\end{cases}
\end{equation}
where $n = 0, \dots, N-1$, together with the condition $\lambda_{n+1} = \lim_{\epsilon\rightarrow 0}u_n(T_{n+1} - \epsilon)$.
The parareal scheme is established by two propagators.
Let $\delta t$ be a fine time step corresponding to the problem (\ref{eq:pde2}).
Suppose that the time-dependent problem is approximated by some appropriate classical discretization schemes.
Let $F$ be a precise integration method based on $\delta t$, often called a fine propagator.
Similarly, let $G$ be a coarse propagator with respect to the broad time step $\Delta T$.
It is seen that the discretization schemes used for such two levels might be different.
Finally, parareal replaces the accurate sequential method by an iteration
\[
\lambda_{n+1}^{k+1} = G(\lambda_n^{k+1}) + F(\lambda_n^k) - G(\lambda_n^k),
\]
which is a predictor-corrector iterative scheme with $n = 0,\dots,N-1$ and $\lambda_0^0 = u_0$.
Additionally, we are obliged to respect two more conditions $\lambda_{n+1}^0=G(\lambda_n^0)$ and
$\lambda_0^{k+1}=\lambda_0^k$ throughout the iterative process.
We note that the time-dependent problems might benefit from such scheme only if the number of iterations is less than $N$.
Following the mathematical model, we arrive at the iterative process below
\begin{alg}
\label{alg:pr}
(Parareal Iterative Algorithm)

\begin{algorithmic}
\STATE $\lambda_0^0 = u_0$
\FOR{$n=0$ \TO $N-1$}
\STATE $\lambda_{n+1}^0 = G(\lambda_n^0)$
\ENDFOR
\STATE{$k=0$}
\WHILE{not convergence}
\STATE{Solve $F(\lambda_n^k)$ in parallel for $n = 0,\dots,N-1$}
\STATE{$\lambda_0^{k+1} = \lambda_0^k$}
\FOR{$n=0$ \TO $N-1$}
\STATE{Solve $G(\lambda_n^{k+1})$}
\STATE{$\lambda_{n+1}^{k+1} =
G(\lambda_n^{k+1}) + F(\lambda_n^k) - G(\lambda_n^k)$}
\ENDFOR
\STATE{$k=k+1$}
\ENDWHILE
\end{algorithmic}
\end{alg}
\noindent Notice that $F$ can be evaluated independently using parallel computers.
Thus, parareal is often preferable than the serial setting for the time-dependent problems.

\subsection{Asynchronous Parareal Scheme}
To derive the corresponding asynchronous scheme, now we present the asynchronous iterative model.
The fundamental asynchronous model (see, e.g., \cite{Bertsekas1989a}) is
\begin{equation}
\label{eq:async1}
x_i^{k+1} =
\begin{cases}
f_i(x_1^{\mu_1^i(k)},\dots,x_p^{\mu_p^i(k)}), & i \in P^k, \\
x_i^k, & i \notin P^k,
\end{cases}
\end{equation}
where $p$ is the number of processor, $P^k \subseteq \{1,\dots,p\}$, $P^k \neq \emptyset$ and $\mu_j^i(k)$ are integers satisfying $0 \le \mu_j^i(k) \le k$.
In addition, the model (\ref{eq:async1}) is subject to the following conditions
\begin{equation}
\label{eq:con1}
\begin{cases}
\underset{k \to +\infty}{\lim}\mu_j^i(k) = +\infty, & i, j \in \{1,\dots,p\}, \\
\abs{\{k \in \mathbb{N} \mid i \in P^k\}} = +\infty, & i \in \{1,\dots,p\},
\end{cases}
\end{equation}
which indicates that each processor reads eventually the latest information for each component and no processor stops updating.
Unfortunately, this model fails to be applied to the two-stage problems.
Therefore, we explore an alternative method which is often called asynchronous iterations with flexible communication, see \cite{Frommer1998}.
To the end we will consider the model
\[
x_i^{k+1} =
\begin{cases}
f_i(x_1^{\mu_1^i(k)},\dots,x_p^{\mu_p^i(k)},x^{\rho^i(k)}), & i \in P^k, \\
x_i^k, & i \notin P^k,
\end{cases}
\]
with
\[
x^{\rho^i(k)} = [x_1^{\rho_1^i(k)},\dots,x_p^{\rho_p^i(k)}],
\]
where $\rho_j^i(k)$ are integers satisfying $0 \le \rho_j^i(k) \le k$ and similarly we have
\begin{equation}
\label{eq:con2}
\underset{k \to +\infty}{\lim}\rho_j^i(k) = +\infty,\quad i, j \in \{1,\dots,p\}.
\end{equation}
In the context of the asynchronous two-stage iterative model, we can eventually derive the asynchronous parareal method.
Consider the following iteration
\[
\lambda_{n+1}^{k+1} =
\begin{cases}
G(\lambda_n^{\mu_n(k)}) + F(\lambda_n^{\rho_n(k)})
- G(\lambda_n^{\rho_n(k)}), & n \in P^k, \\
\lambda_{n+1}^k, & n \notin P^k,
\end{cases}
\]
where $\lambda_0^{k+1} = \lambda_0^k$, $P^k \subseteq \{1,\dots,p\}$ and $P^k \neq \emptyset$, with the similar conditions
\[
0 \le \mu_j^i(k) \le k+1,\quad 0 \le \rho_j^i(k) \le k.
\]
We supposed that the essential assumptions (\ref{eq:con1}) and (\ref{eq:con2}) are satisfied.

\subsection{Asynchronous Parareal Scheme}

To derive the corresponding asynchronous scheme, now we present the asynchronous iterative model.
The fundamental asynchronous model (see, e.g., \cite{Bertsekas1989a}) is
\begin{equation}
\label{eq:async1}
x_i^{k+1} =
\begin{cases}
f_i(x_1^{\mu_1^i(k)},\dots,x_p^{\mu_p^i(k)}), & i \in P^k, \\
x_i^k, & i \notin P^k,
\end{cases}
\end{equation}
where $p$ is the number of processor, $P^k \subseteq \{1,\dots,p\}$, $P^k \neq \emptyset$ and $\mu_j^i(k)$ are integers satisfying $0 \le \mu_j^i(k) \le k$.
In addition, the model (\ref{eq:async1}) is subject to the following conditions
\begin{equation}
\label{eq:con1}
\begin{cases}
\underset{k \to +\infty}{\lim}\mu_j^i(k) = +\infty, & i, j \in \{1,\dots,p\}, \\
\abs{\{k \in \mathbb{N} \mid i \in P^k\}} = +\infty, & i \in \{1,\dots,p\},
\end{cases}
\end{equation}
which indicates that each processor reads eventually the latest information for each component and no processor stops updating.
Unfortunately, this model fails to be applied to the two-stage problems.
Therefore, we explore an alternative method which is often called asynchronous iterations with flexible communication, see \cite{Frommer1998}.
To the end we will consider the model
\[
x_i^{k+1} =
\begin{cases}
f_i(x_1^{\mu_1^i(k)},\dots,x_p^{\mu_p^i(k)},x^{\rho^i(k)}), & i \in P^k, \\
x_i^k, & i \notin P^k,
\end{cases}
\]
with
\[
x^{\rho^i(k)} = [x_1^{\rho_1^i(k)},\dots,x_p^{\rho_p^i(k)}],
\]
where $\rho_j^i(k)$ are integers satisfying $0 \le \rho_j^i(k) \le k$ and similarly we have
\begin{equation}
\label{eq:con2}
\underset{k \to +\infty}{\lim}\rho_j^i(k) = +\infty,\quad i, j \in \{1,\dots,p\}.
\end{equation}
In the context of the asynchronous two-stage iterative model, we can eventually derive the asynchronous parareal method.
Consider the following iteration
\[
\lambda_{n+1}^{k+1} =
\begin{cases}
G(\lambda_n^{\mu_n(k)}) + F(\lambda_n^{\rho_n(k)})
- G(\lambda_n^{\rho_n(k)}), & n \in P^k, \\
\lambda_{n+1}^k, & n \notin P^k,
\end{cases}
\]
where $\lambda_0^{k+1} = \lambda_0^k$, $P^k \subseteq \{1,\dots,p\}$ and $P^k \neq \emptyset$, with the similar conditions
\[
0 \le \mu_j^i(k) \le k+1,\quad 0 \le \rho_j^i(k) \le k.
\]
We supposed that the essential assumptions (\ref{eq:con1}) and (\ref{eq:con2}) are satisfied.

\section{Formalization of the Option Pricing Model}
\label{sec:3}

We now turn to the study of the option pricing problem that is expressed by appropriate evolution models.
Accordingly, they apply to the above described scenario.
To the end we will concentrate on the European call option, which can be exercised only on the expiration date, using the Black-Scholes model as following
\begin{equation}
\label{eq:bs}
\frac{\partial V}{\partial t} +  rS\frac{\partial V}{\partial S} +
\frac{1}{2}\sigma^2S^2\frac{\partial^2 V}{\partial S^2} = rV,
\end{equation}
where $V$ is the option price as a function of underlying asset price $S$ and time $t$.
The parameters are volatility $\sigma$ and risk-free interest rate $r$.
Furthermore, the Black-Scholes equation is backward parabolic satisfying
\[
\begin{cases}
V(S,T) = \max(S-E, 0), & S \ge 0, \\
V(0,t) = 0, & t \ge 0, \\
V(S,t) \sim S - Ee^{-r(T-t)}\ as\ S \rightarrow +\infty, & t \ge 0,
\end{cases}
\]
where the first formula is a final condition, $T$ is the time to maturity and $E$ is the exercise price.
We are particularly interested in the solution of its discretization.
However, the irregular form poses some difficulties for the solution.
Therefore, we now try to transform the formula (\ref{eq:bs}) to a simplest case.

A well known special case is the heat equation, which leads to a really simple parabola.
Thus, the transformation from the original to the latter can simplify the solving process.
Therefore, a change of variables is
\[
S = Ee^x,\quad t = T - \frac{2\tau}{\sigma^2},\quad V = Ev(x, \tau),
\]
substituting into the Black-Scholes equation (\ref{eq:bs}) gives
\[
\frac{\partial v}{\partial\tau} = (\kappa-1)\frac{\partial v}{\partial x}
+ \frac{\partial^2 v}{\partial x^2} - \kappa v,
\]
where $\kappa = \frac{2r}{\sigma^2}$. Setting
\begin{equation}
\label{eq:ab}
\alpha = \frac{1}{2}(\kappa-1),\quad \beta = \frac{1}{2}(\kappa+1),
\end{equation}
yields
\[
v = e^{-\alpha x-\beta^2\tau}u(x,\tau),
\]
then gives the heat equation in an infinite interval
\begin{equation}
\label{eq:heat}
\frac{\partial u}{\partial \tau} = \frac{\partial^2 u}{\partial x^2},\quad
x\in\mathbb{R},\ \tau\in[0, \frac{T\sigma^2}{2}].
\end{equation}
heat equation is forward parabolic as opposed to the original model, with the corresponding conditions
\[
\begin{cases}
u(x,0) = \max(e^{\beta x}-e^{\alpha x}, 0),
& x\in\mathbb{R}, \\
u(x,\tau) \sim 0\ as\ x \rightarrow \pm\infty, & \tau\in[0, \frac{T\sigma^2}{2}].
\end{cases}
\]
We notice here that equation (\ref{eq:heat}) corresponds to the initial problem (\ref{eq:pde1}) with $f = 0$ and $\mathcal{L} = -\frac{\partial^2}{\partial x^2}$.
Hence, the parareal schemes can be used to approximate the solution.

We assume that the equation (\ref{eq:heat}) is discretized in time using a two-stage scheme.
Generally, the same discretization approach as coarse solver or a method of higher order can be applied to the fine grid propagator.
Nevertheless, the spatial discretization is not straightforward enough.
We give the precise boundaries as following instead of limits to infinity
\[
\begin{cases}
x^- = \min(x_0, 0) - \log(4), \\
x^+ = \max(x_0, 0) + \log(4),
\end{cases}
\]
such that
\[
\begin{cases}
u(x^-, \tau) = 0, & \tau\in[0, \frac{T\sigma^2}{2}], \\
u(x^+, \tau) = e^{\beta x^+ + \beta^2\tau} - e^{\alpha x^+ + \alpha^2\tau},
& \tau\in[0, \frac{T\sigma^2}{2}],
\end{cases}
\]
where $\alpha$ and $\beta$ are defined by (\ref{eq:ab}).

\section{Experimental Results}
\label{sec:4}

In this section we present numerical results for the above asynchronous iterative scheme in order to solve the Black-Scholes equation.
Let us give the volatility $\sigma=0.2$ and the risk-free interest rate $r=0.05$, which can be considered as constants since they have little relation with the asynchronous properties.
Another intuitive configuration is to set $\delta t$ equals one day, while taking into account $T=1$ representing one year.
However, such assignment seems inflexible and yields a somewhat complicated constant.
Hence, we keep $\delta t=0.001$ fixed and vary $\Delta T$ to simulate the different remaining life of the financial instrument.
Finally, the spatial discretization is implemented by the finite difference method with suitable equal sub-intervals.

The asynchronous implementation requests some advanced capabilities.
We conduct our tests based on JACK (an asynchronous communication kernel) library \cite{Magoules2017b}, which provides high-level supports for both synchronous and asynchronous iterative algorithms, developed on the top of MPI library.
Then, mathematical operations and linear systems solvers are implemented by Alinea (An Advanced Linear Algebra) library \cite{Magoules2015a}
The Alinea library is implemented in C++, MPI, CUDA and OpenCL.
For both central processing unit and graphic processing unit devices, there are different matrix storage formats, and real and complex arithmetics in single- and double-precision.  
It includes several linear algebra operations \cite{Ahamed2016c} and numerous algorithms for solving linear systems such as iterative methods \cite{Magoules2015b}, \cite{Ahamed2016d},  \cite{Magoules2015c}, together with some energy consumption optimization \cite{Magoules2016b}, and domain decomposition methods in space \cite{Magoules2016c}. 
We carry out two sets of experiments on a SGI ICE X cluster connected with InfiniBand (56 Gbit/s).
Each computing node consists of two Intel Xeon E5-2670 v3 2.30 GHz CPUs.
Furthermore, the version of the MPI library used is SGI-MPT 2.14.

The first is performed by changing the length of $\Delta T$ on which we simulate the different time to maturity.
The backward Euler method is chosen for both the coarse and the fine propagator.
Given the initial stoke price $S=15$ and the strike price $E=20$, table \ref{tab:res1} illustrates the some average results for 16 cores and 250 sub-intervals, where $V_a$ is the approximate option prices, $V_e$ is the exact price, $\epsilon_a$ is the absolute error and $\epsilon_r$ is the relative error with $\epsilon_r = \frac{\epsilon_a}{V_e}$.
\begin{table}[!t]
\caption{Results of the Asynchronous Parareal Scheme for 16 Cores, with
Approximate Option Prices $V_a$, Exact Option Prices $V_e$, Absolute Error
$\epsilon_a$ and Relative Error $\epsilon_r$ ($\delta t=0.001$, $S=15$,
$E=20$, $m=250$)}
\label{tab:res1}
\centering
\begin{tabular}{|c||c|c|c|c|c|}
\hline
$\Delta T$ & $V_a$ & $V_e$ & $\epsilon_a$ & $\epsilon_r$ & Time \\
\hline
0.1 & 0.4857 & 0.4853 & 0.0004 & 0.0008 & 0.958 \\
0.2 & 1.3950 & 1.3947 & 0.0003 & 0.0002 & 2.012 \\
0.3 & 2.3145 & 2.3140 & 0.0005 & 0.0002 & 3.014 \\
0.4 & 3.1932 & 3.1925 & 0.0007 & 0.0002 & 4.091 \\
0.5 & 4.0212 & 4.0203 & 0.0009 & 0.0002 & 5.041 \\
0.6 & 4.7972 & 4.7961 & 0.0011 & 0.0002 & 6.084 \\
0.7 & 5.5226 & 5.5213 & 0.0013 & 0.0002 & 6.873 \\
0.8 & 6.1995 & 6.1981 & 0.0014 & 0.0002 & 7.925 \\
0.9 & 6.8306 & 6.8291 & 0.0015 & 0.0002 & 8.526 \\
1.0 & 7.4184 & 7.4169 & 0.0015 & 0.0002 & 9.969 \\
\hline
\end{tabular}
\end{table}
Table \ref{tab:res2} reports the similar results with $S=25$ and $E=30$.
\begin{table}[!t]
\caption{Results of the Asynchronous Parareal Scheme for 16 Cores
($\delta t=0.001$, $S=25$, $E=30$, $m=250$)}
\label{tab:res2}
\centering
\begin{tabular}{|c||c|c|c|c|c|}
\hline
$\Delta T$ & $V_a$ & $V_e$ & $\epsilon_a$ & $\epsilon_r$ & Time \\
\hline
0.1 & 1.5204 & 1.5179 & 0.0025 & 0.0016 & 0.969 \\
0.2 & 3.3163 & 3.3141 & 0.0022 & 0.0007 & 1.972 \\
0.3 & 4.9525 & 4.9504 & 0.0021 & 0.0004 & 3.022 \\
0.4 & 6.4497 & 6.4476 & 0.0021 & 0.0003 & 3.817 \\
0.5 & 7.8264 & 7.8241 & 0.0023 & 0.0003 & 5.221 \\
0.6 & 9.0963 & 9.0939 & 0.0024 & 0.0003 & 6.108 \\
0.7 & 10.2699 & 10.2673 & 0.0026 & 0.0003 & 6.953 \\
0.8 & 11.3557 & 11.3530 & 0.0027 & 0.0002 & 7.731 \\
0.9 & 12.3611 & 12.3584 & 0.0027 & 0.0002 & 8.871 \\
1.0 & 13.2923 & 13.2898 & 0.0025 & 0.0002 & 10.396 \\
\hline
\end{tabular}
\end{table}
The absolute errors and the relative errors are small enough, which leads to the conclusion that the asynchronous iterations converge to the exact solution of the Black-Scholes equation.
Therefore, the fact above illustrates that the asynchronous parareal scheme is suitable to the option pricing problem, formalized by the Black-Scholes model.
Notice that $T=16$ when $\Delta T=1$, which is beyond the scope of common requirements, even in a more general context.

The second is a comparison of synchronous and asynchronous parareal scheme.
We choose $S=25$, $E=30$, $\Delta T=0.1$ and 150 sub-intervals.
Table \ref{tab:res3} reports the average experimental results, where $N$ is the number of processor cores.
\begin{table}[!t]
\caption{Comparison of Synchronous and Asynchronous Parareal Scheme
($S = 25$, $E = 30$, $\delta t = 0.001$, $\Delta T = 0.1$, $m = 150$)}
\label{tab:res3}
\centering
\begin{tabular}{|c||c|c|c|c|c|c|c|}
\hline
& \multicolumn{2}{c|}{Synchronous}
& \multicolumn{4}{c|}{Asynchronous} \\
\hline
$N$ & Iter. & Time & \multicolumn{3}{c|}{Iter. (min, max, mean)} & Time \\
\hline
16 & 11 & 0.620 & 22 & 30 & 26 & 0.490 \\
32 & 11 & 0.781 & 30 & 47 & 40 & 0.677 \\
64 & 11 & 0.971 & 44 & 77 & 60 & 0.947 \\
\hline
\end{tabular}
\end{table}
The results illustrate that the number of iterations of the synchronous version remains the same throughout the test, whereas the asynchronous scheme requires further iterations with the increase of cores.
Let us mention here that the context is not distributed, which rejects the possibility of large
communication delays.
We notice that the asynchronous version is faster.

\section{Conclusions}
\label{sec:5}

In this paper we show the convergence performance of an original asynchronous extension of the parareal scheme applied to the Black-Scholes equation, and illustrate some cases that the asynchronous method takes on superiority than its synchronous counterpart.
The asynchronous iterative methods performs well even if there exists large communication delays or the computation bottlenecks in the systems, so long as all the processors keep updating using eventually the latest values.
We illustrate the excellent precision and the conditional efficiency for the target approach.
A theoretical and comprehensive analysis is under investigation by the authors to explain the results.

\bibliography{ref}
\bibliographystyle{abbrv}

\end{document}